\documentclass[twocolumn,showpacs,prl]{revtex4}
\usepackage{amsmath}
\usepackage{graphicx}
\usepackage{dcolumn}
\usepackage{bm}

\begin{document}

\title{Fluctuation conductivity of thin films and nanowires near a
parallel-field-tuned superconducting quantum phase transition}
\author{A.~V.~Lopatin$^{1}$,
N.~Shah$^{2,3}$ and V.~M.~Vinokur$^{1}$ }
\affiliation{$^{1}$
Materials Science Division, Argonne National
Laboratory, Argonne, Illinois 60439, USA.\\
$^{2}$ Festk\"orpertheorie, Paul Scherrer Institut, CH-5232
Villigen PSI, Switzerland.\\
$^{3}$ Theoretische Physik, ETH-H\"onggerberg, CH-8093 Z\"urich,
Switzerland.}
\date{\today}

\begin{abstract}
We calculate the fluctuation correction to the normal state
conductivity in the vicinity of a quantum phase transition from a
superconducting to normal state, induced by applying a magnetic
field parallel to a dirty thin film or a nanowire with thickness
smaller than the superconducting coherence length. We find that at
 zero temperature, where the correction comes purely from
quantum fluctuations, the positive \textquotedblleft
 Aslamazov-Larkin\textquotedblright\ contribution, the negative
\textquotedblleft density of states\textquotedblright\
contribution, and the \textquotedblleft
Maki-Thompson\textquotedblright\ interference contribution, are
all of the same order and the total correction is negative.
Further we show that based on how the quantum critical point is
approached, there are three regimes that show different
temperature and field dependencies which should be experimentally
accessible.
\end{abstract}

\maketitle

The high interest in the physics of a quantum critical point (QCP)
is motivated by the explosive growth of its experimental
realizations, driving, in turn, a quest for theoretical
understanding of quantum phase transitions (QPT)\cite{Sachdev_book}.
The challenge for theory is two fold:
First, it is desirable to identify experimentally accessible
systems experiencing QPT. Second, these systems should allow for a
systematic and comprehensive theoretical description. In this
respect, the QCP realized in dirty superconductors of reduced
dimensions\cite{Gantmakher,Liu01} under application of magnetic field represents
an exemplarily controllable
system allowing complete experimental exploration of the vicinity
of the critical point on the one hand and systematic theoretical
study on the other.

In this letter we present a full systematic investigation of
fluctuation corrections\cite{Craven73,LarkinV} to the normal state conductivity of a thin
wire or a thin film in the vicinity of a QPT
from superconducting to normal state, induced by an applied
magnetic field.  We consider a thin wire (or thin film) of
diameter $d$ (or thickness $t$ for the film) much smaller than the
superconducting coherence length $\xi $. The magnetic field $H$ is
applied along the wire (or parallel to the film) and can be
parameterized completely by a scalar pair-breaking parameter
\cite{Larkin65}(similar to the effect of paramagnetic
impurities\cite{Abrikosov61})
\begin{equation}
\alpha =\left\{
\begin{array}{ccc}
D\,(eHd/2c)^{2}/4 &  & \text{wire } \\
D\,(eHt/c)^{2}/6 &  & \text{film }
\end{array}
\right.  \label{alpha}
\end{equation}
where $D$ is the diffusion coefficient in the bulk sample
\cite{Zeeman}. The superconducting critical temperature $T_c$ is
related to $\alpha$ (see Fig. 1)  via the standard equation
\begin{equation}
\ln (T_{c}/T_{c0})=\psi (1/2)-\psi (1/2+\alpha /2\pi T_{c}),
\label{Tc_vs_H}
\end{equation}
where $\psi $ is the digamma function and $ T_{c0}\equiv
T_{c}(H=0)$ is the critical temperature in the absence of a
magnetic field. One can immediately see that at $T=0$ the
superconductivity is destroyed at $\alpha _{c0}\equiv \alpha
_{c}(T=0)=\pi T_{c0}/2\gamma $, where $ \ln\gamma \approx 0.577$
is the Euler constant. Realization of the superconducting QCP in a
superconductor with paramagnetic impurities was first suggested in
the work of Ref.\cite{RamazashviliC97} (see also
Ref.\cite{MineevS01}). Since in our case the depairing parameter
$\alpha$ depends on $H$, it allows for a well controlled
exploration of the QPT and its vicinity by varying the applied
magnetic field.

\begin{figure}[tbp]
\resizebox{.46\textwidth}{!}{\includegraphics{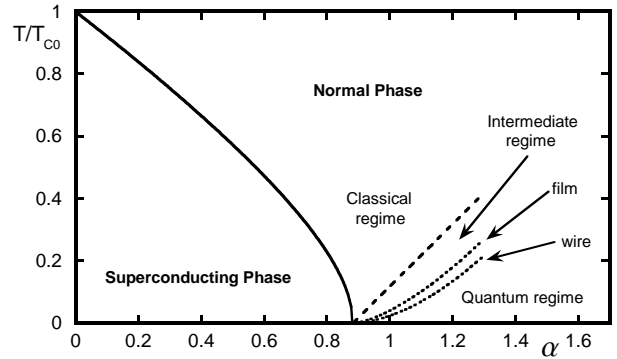}}
\vspace{-0.2cm} \caption{ Phase diagram of a superconducting
nanowire (thin film) in a parallel magnetic field (parametrized by $\alpha$).
The boundary between the quantum and the intermediate regime, shown by a dotted line
is different for a wire and film as marked in the figure while all other boundaries coincide.}
\label{fig:1}
\end{figure}

Our main result for the fluctuation correction to the normal state
conductivity can be presented as
\begin{equation}
\delta \sigma (\alpha ,T)=\delta \sigma _{0}(\alpha )+\delta
\sigma _{\scriptscriptstyle T}(\alpha ,T),  \label{result}
\end{equation}
where $\delta \sigma _{0}(\alpha )$  and $\delta \sigma_T (\alpha
,T)$  are the zero temperature and finite temperature
contributions, respectively. The zero temperature correction is
given by
\begin{equation}
\delta \sigma _{0}(\alpha)=-{\frac{{2De^{2}}}{{\pi d(2+d)}}}\int
{\frac{{ d^{d}q}}{{(2\pi )^{d}}}}{\frac{{(Dq^{2})^{2}}}{{\alpha
_{q}^{3}\,\ln (\alpha _{q}/\alpha _{c})}},}  \label{final_T=0}
\end{equation}
where $\alpha_{q}\equiv \alpha +Dq^{2}/2 $ and $d=1,2$ corresponds
to a wire and film respectively. Its magnitude decreases
monotonically with increasing field; this leads to a {\it
negative} magnetoresistance. Note that a negative
magnetoresistance was found also in granular
superconductors\cite{Beloborodov} and in thin films in {\it
perpendicular} magnetic field\cite{Galitskii01}. Shown in Fig.2
are plots of the dimensionless correction
\begin{equation}
\delta \bar{\sigma}_{0}(\alpha)=e^{-2}\times \left\{
\begin{array}{cccc}
(D/ \alpha_{c0}  )^{-1/2} \, \delta \sigma _{0}(\alpha )\; &  &  & \text{wire} \\
\delta \sigma _{0}(\alpha ) &  &  & \; \text{film.\; }
\end{array}
\right.  \label{zero_temp_result}
\end{equation}
 When expanded around the QCP, the dimensionless conductivity close to the QCP is given by
\begin{equation}
\delta \bar{\sigma}_{0}(\alpha )=\delta \bar{\sigma}_{0}(\alpha =\alpha
_{c0})+\,b(\alpha -\alpha _{c0})/\alpha _{c0},  \label{expansion}
\end{equation}
with the numerical coefficient $b=0.386$ and $b=0.070$ for a wire
and film, respectively.  Thus the zero temperature critical
exponent is 1 for both $d=1,2$, at least within the perturbation
theory.

In the vicinity of the QCP ($T,{\alpha -\alpha _{c}\ll }\alpha
_{c0}$ ), the field dependence of $\delta \sigma_{T}(\alpha ,T)$ turns
out to be more singular than that of $\delta \sigma _{0}(\alpha )$
and for $T>\alpha -\alpha _{c}$ its leading term is given by
\begin{equation}
\delta \sigma _{\scriptscriptstyle T}(\alpha ,T)=e^{2}\times \left\{
\begin{array}{cccc}
{\frac{{\sqrt{D}T}}{{4\sqrt{2}(\alpha -\alpha _{c})^{3/2}}}} &  &  & \text{%
wire} \\
{\frac{{T}}{{4\pi (\alpha -\alpha _{c})}}} &  &  & \text{film,}
\end{array}
\right.   \label{classical}
\end{equation}
while for $T<\alpha -\alpha _{c}$ it is
\begin{equation}
\label{intermediate}
\delta \sigma _{\scriptscriptstyle T}(\alpha ,T)=e^{2}\times
\left\{
\begin{array}{cccc}
{\frac{{\pi \sqrt{D}T^{2}}}{{12\sqrt{2}(\alpha -\alpha _{c})^{5/2}}}} &  &
& \text{wire} \\
{\frac{{T^{2}}}{{18(\alpha -\alpha _{c})^{2}}}} &  &  & \text{film.}
\end{array}
\right.
\end{equation}
The contributions $\delta \sigma _{\scriptscriptstyle T}$ and
$\delta \sigma _{0}$ become comparable at
\begin{equation}
T\equiv T_{0}(\alpha )\sim \left\{
\begin{array}{cccc}
(\alpha -\alpha _{c0})^{7/4}/\,\alpha _{c0}^{3/4} &  &  & \text{wire} \\
 (\alpha -\alpha _{c0})^{3/2}/\,\alpha _{c0}^{1/2} &  &  & \text{film.}
\end{array}
\right.
\end{equation}
The key point is that the behavior of the fluctuation corrections
to the conductivity depends on the way one approaches the QCP and
we can identify three regimes in the vicinity of the QCP that show
qualitatively different
behaviors as illustrated in Fig. 1. There is a ``classical'' regime for $%
T>\alpha -\alpha _{c}$, where the correction is given by Eq.(\ref{classical}%
); an ``intermediate'' regime for $\alpha -\alpha _{c}>T>T_{0}(\alpha ),$
where the correction behaves according to Eq.(\ref{intermediate} ); and a
``quantum'' regime for $T_{0}(\alpha )>T$ where the behavior crosses over to
an essentially zero-temperature-like behavior which is not singular and
almost temperature independent with the fluctuation correction dominated by $%
\delta \sigma _{0}$ as given by Eqs.(\ref{final_T=0},\ref{zero_temp_result},\ref{expansion}).

\begin{figure}[tbp]
\resizebox{.46\textwidth}{!}{\includegraphics{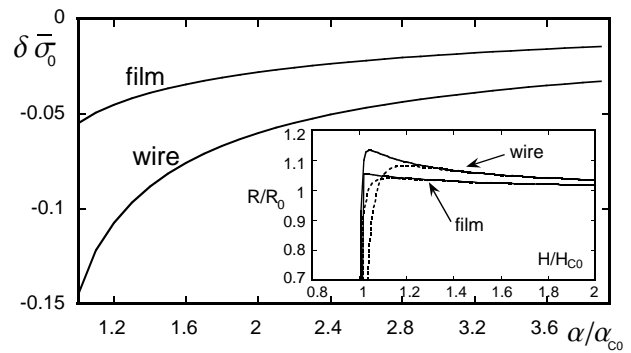}}
\vspace{0.1cm} \caption{ Dimensionless zero temperature
fluctuation conductivity correction(\ref{zero_temp_result}) as a
function of depairing parameter $\alpha .$ Insert shows dependence
of the resistivity on the magnetic filed for temperatures
$T/\alpha_{c0} = 0.01$ (upper curves) and  $T / \alpha_{c0} = 0.1$
(lower curves). Resistivity is normalized to the high temperature
resistivity $R_0$ while the magnetic field is normalized to the
critical filed $H_{c0}$  at  $T=0.$ }\label{fig:2}
\end{figure}

Since in the quantum region the correction to conductivity is
negative whereas in the classical and intermediate region it is positive, we predict a
non-monotonic behavior of the resistivity as a function of the
magnetic field at finite temperature. The corresponding plots are
shown in the insert of Fig. 2 for nanowires and thin films.
Such behavior was indeed observed in experiments on amorphous
thin films\cite{Gantmakher}, while for nanowires, to the best of
our knowledge, it was not reported yet.

Our approach is based on the diagrammatic perturbation theory.
Note that the technique using the time-dependent Ginzburg-Landau
formalism that was adapted in Refs.\cite{RamazashviliC97,MineevS01},
accounts only for the direct ``Aslamazov-Larkin'' (AL) type of contribution\cite{AslamazovL68}
to the fluctuation
conductivity that comes from the charge transfer via fluctuating
Cooper pairs, but misses the zero-temperature contribution to the
correction. On the contrary, our approach, takes care of all the contributions including the
``density of states'' (DOS) part resulting from the reduction of the normal
single-electron density
of states at the Fermi level, and the more indirect
``Maki-Thompson''\ (MT) interference
contribution\cite{Maki68,Thompson70}.
At zero temperature, where the correction comes purely from quantum fluctuations,
these turn out to be of the same order as the AL contribution.

To derive our main results, we carry out a microscopic calculation
within the standard framework of temperature diagrammatic
technique in a disordered electron system\cite{AGD, AVR,
Altshuler80} in the diffusive limit (inverse mean free time $\tau
^{-1}\gg T,\alpha $). Since we choose $ d,t<\xi $ and parameterize
the effect of field by $\alpha $ as defined by Eq.( \ref{alpha}),
the problem effectively becomes one and two dimensional for a wire
and film, respectively. The main building block of the
diagrammatic technique in the presence of the BCS interaction is
the so-called ``Cooperon'', the ladder diagram that describes
coherent scattering by impurities in the particle-particle
channel. In the presence of a parallel magnetic field it is given
by
\begin{equation}
C(\omega _{1},\omega _{2},q)={\frac{{2\pi \nu \theta (-\omega _{1}\omega
_{2})}}{{|\omega _{1}-\omega _{2}|+Dq^{2}+2\alpha }}},  \label{Cooperon}
\end{equation}
where ${\theta }$ is the step function,  $\nu $ is the
density of states at the Fermi surface, and  $\mathbf{q}$ is the
momentum in the effective dimension. Using Eq.(\ref{Cooperon}) in
a standard way, one obtains the ``fluctuation propagator'' i.e the
impurity-averaged sum over the ladder diagrams corresponding to
the electron-electron interaction in the Cooper channel,
\begin{equation}
K^{-1}(\Omega ,q)=\ln \left( {\frac{T}{{T_{c}^{0}}}}\right) -\psi
\left( {\frac{1}{2}}\right) +\psi \left( {\frac{1}{2}}+{\frac{{\alpha
_{q}+|\Omega |/2}}{{2\pi T}}}\right),  \label{Pair_Correlator}
\end{equation}
 where $\Omega$ is the bosonic Matsubara
frequency. The pole of Eq.(\ref{Pair_Correlator}) defines the
boundary between the superconducting and normal phases given by
Eq.(\ref{Tc_vs_H}). At low temperatures, $T\ll \alpha _{c0},$ the
fluctuation propagator reduces to
\begin{equation}
K^{-1}(\Omega ,q)=\ln [(\alpha _{q}+|\Omega )|/2)/\alpha _{c}(T)].
\end{equation}

The fluctuation correction to the conductivity is obtained as usual from the Kubo
formalism with the appropriate analytic continuation. The standard
set of diagrams constituting the AL, DOS and MT\ contributions is shown in Fig. 3.

 The AL contribution (Fig. 3a) can be expressed as a sum of two
terms: $\delta \sigma ^{AL} =\delta \sigma _{1}^{AL}+\delta \sigma
_{2}^{AL}$ with
\begin{eqnarray}
{\frac{{\delta \sigma _{1}^{AL}}}{{e^{2}}}} &=& {\frac{{D^{2}}}{{2\pi T d}}}%
\int {\frac{{d^{d}q}}{{(2\pi )^{d}}}\frac{{d\Omega }}{{{\rm sh}^{2}   {\frac{{\Omega }%
}{{2T}}}}}} \Bigl[
\mathrm{Im} \{ K(-i\Omega ,q) \, \Gamma_{\Omega \mathbf{q}}^{2}  \}     \nonumber  \\
  \times  & \mathrm{Im}& \{ K(-i\Omega ,q )\}+ \left(\mathrm{Im} \{
  K(-i\Omega ,q) \, \Gamma_{\Omega \mathbf{q}}    \}\right) ^{2} \Bigr], \\
  \vspace{0.2cm}
{\frac{\delta {\sigma _{2}^{AL}}}{{e^{2}}}}
&=&{\frac{{D^{2}i}}{{8\pi ^{4}T^{3}}}}\int \frac{{d^{d}q \,
d\Omega }}{{(2\pi )^{d}}}  \;q_{x}^{2}\psi
^{\prime }\left( {\frac{1}{2}}+{\frac{\alpha_{q} - i\Omega/2 }{{2\pi T}}}%
\right)  \nonumber \\
&&\times \psi ^{\prime \prime }\left( {\frac{1}{2}}+{\frac{\alpha_{q}{%
-i\Omega/2 }}{{2\pi T}}}\right) K^{2}(-i\Omega,q),
\end{eqnarray}
where $ \Gamma _{\Omega \mathbf{q}} \equiv (\mathbf{q}/2\pi T)
\psi ^{\prime }[ 1/2+ (\alpha_q-i\Omega /2)/2\pi T]. $
 The contribution $\delta \sigma _{1}^{AL}$ appears only at finite
temperature while $\delta {\sigma _{2}^{AL}}$ contains the
contribution that survives even at $T=0$.
  Note that $\delta {\sigma _{2}^{AL}}$ that describes
the quantum regime, results from differentiating $\Gamma _{\mathbf{q}%
}(-i\Omega )$ w.r.t $\Omega $ and is missed within the usual static
approximation ( $\Omega=0$).
\begin{figure}[tbp]
\resizebox{.46\textwidth}{!}{\includegraphics{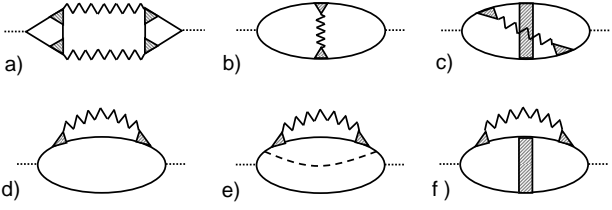}} \vspace{0.1cm}
\caption{ Diagrams for the fluctuation conductivity:
a) Aslamazov-Larkin diagram, b)-c) Maki-Thompson diagrams, d)-e) density of
states diagrams. Full line stands for the disorder averaged normal state Green's function,
 wavy line for the fluctuation propagator $K$, the shaded rectangle for the Cooperon $C$ and
 shaded triangle for the vertex $C/2\pi\nu\tau$.}
\label{fig:3}
\end{figure}

The MT correction is given by $ \delta \sigma ^{MT} =\delta \sigma
_{1}^{MT}+\delta \sigma _{2}^{MT} $ where
\begin{eqnarray}
\frac{{\delta \sigma _{1}^{MT}}}{{e^{2}}} & = &{\frac{{D}}{{2\pi }}}  \int \frac{%
{d^{d}q  \,  d\omega d\Omega  }}{{(2\pi )^{d}}}
\text{th}{\frac{{\omega }}{{2T}}} K(-i\Omega,q)
  \left[
{\frac{\text{cth}{{\frac{\Omega }{{2T}}} }}{{(\alpha _{q}-i\omega
_{+})^{3}}}}   \right.  \nonumber  \\
 && \hspace{2.5cm}
  \left. +{\frac{{i} }{2T{\rm {sh} ^{2}{\frac{{\Omega }}{{2T}}}}}%
\frac{{1}}{\alpha _{q}^{2}{+\omega _{+}^{2}}}} \right], \\
\frac{{\delta \sigma _{2}^{MT}}}{{e^{2}}} &=&-{\frac{{3D^{2}}}
{{2\pi }}}\int \frac{{d^{d}q \, d\omega d \Omega    }}{{(2\pi
)^{d}}} \,  {{ \text{th} {\omega \over{2T}} }\over {\text{th}
{\Omega \over{2T}} }} \,  {{ q_{x}^{2}   K(-i\Omega,q)  } \over {
 (\alpha _{q}-i\omega _{+})^{4}}},
\end{eqnarray}
with ${\omega _{+}\equiv \omega +\Omega /2}$. For low
temperatures, the contributions $\delta \sigma _{1}^{MT}$ (Fig.
3b)\ and $\delta \sigma _{2}^{MT}$ (Fig. 3c) are of the same order
and both need to be taken into account. On the contrary, at high
temperatures, the diagram (Fig. 3c) having an extra Cooperon
propagator is of a lower order.

The DOS fluctuation correction is given by the expression $\delta
\sigma ^{DOS} =\delta \sigma _{1}^{DOS}+\delta \sigma _{2}^{DOS}$
where
\begin{eqnarray}
 \nonumber {\frac{{\delta \sigma _{1}^{DOS}}}{{e^{2}}}}
&=& {\frac{{D}}{{2\pi }}}\int
\frac{{d^{d}q \, d\omega d\Omega    }}{{(2\pi )^{d}}} \text{th}{\frac{{\omega }}{{2T}%
}} K(-i\Omega ,q)
   \left[ {\frac{{  {\rm cth}  {\frac{\Omega }{{2T}}} }} {{(\alpha_{q}-i\omega _{+})^{3}}}} \right.
 \\
&&
\left. \hspace{1.8cm} -{\frac{{i} }{2T{ {\rm sh}^{2} {\frac{{\Omega  }}{{2T}}}}}}Re{%
\frac{{1}}{{(\alpha _{q}-i\omega _{+})^{2}}}}\right] ,
 \\
\nonumber{\frac{{\delta \sigma _{2}^{DOS}}}{{e^{2}}}} &=&{\frac{{D^{2}}}{{4\pi }}}%
\int \frac{{d^{d}q \, d\omega d\Omega }}{{(2\pi )^{d}}}  q_{x}^{2}\text{th}{\frac{{%
\omega }}{{2T}}} K(-i\Omega ,q)
  \\
& \times & \left[ {\frac{{ - 3}\text{cth}{{\frac{\Omega }{{2T}}}%
}}{{(\alpha _{q}-i\omega _{+})^{4}}}}
 +{\frac{{i}  }{T{ {\rm sh} ^{2}{\frac{{\Omega }}{{2T}}}}}}%
Re{\frac{{1}}{{(\alpha _{q}-i\omega _{+})^{3}}}}\right] .
\end{eqnarray}
 The correction $\delta \sigma _{1}^{DOS}$corresponds to diagrams shown in Fig.
3d, 3e while $\delta \sigma _{2}^{DOS}$ is given by diagrams from
 Fig. 3f having an extra Cooperon propagator.

Equipped with the analytical expressions for all the necessary
diagrams, we now proceed to their evaluation beginning with the
zero temperature case. One can see that apart from $\delta {\sigma
_{2}^{AL}}$, only the terms in the DOS\ and MT\ contributions that
have a cth$(\Omega )$ survive at $T=0.$ Integrating over frequency
$\Omega $ by parts, the $T=0$ answers for all contributions can be
expressed in terms of just two functions defined by
\begin{eqnarray}
F_{1} &=&{\frac{{2D}}{{\pi d}}}\int {\frac{{d^{d}q}}{{(2\pi )^{d}}}}{\frac{{%
Dq^{2}}}{{\alpha _{q}^{2}}}}K(\Omega =0,q)  \label{F1} \\
F_{2} &=&-{\frac{{2D}}{{3\pi (2+d)d}}}\int {\frac{{d^{d}q}}{{(2\pi )^{d}}}}{%
\frac{{(Dq^{2})^{2}K_{\alpha }^{\prime }(\Omega =0,q)}}{{\alpha _{q}^{2}}},}
\label{F2}
\end{eqnarray}
to yield $\delta \sigma ^{AL}=6F_{2},\;\delta \sigma
^{DOS}=-(F_{1}+3F_{2})/2,\;$and $\delta \sigma ^{MT}=-3F_{2}$
\cite{remark}. Since the functions $F_{1}$ and $F_{2}$ are
positive, we find that the AL correction is positive and the DOS
correction is negative as expected while the MT correction which
doesn't have a prescribed sign is negative. Finally the
fluctuation correction to the normal state conductivity given by
$\delta \sigma _{0}^{total}=-(F_{1}-3F_{2})/2,$ after integrating
by parts results in Eq.(\ref{final_T=0}).

At nonzero temperatures, the terms having a factor $%
1/$sh$(\Omega /2T)$ need to be included. One finds that the
leading contribution close to the QCP\ comes from $\delta {\sigma
_{1}^{AL}}$ and the evaluation of this term in different regimes
leads to the results Eq.(\ref{classical},\ref{intermediate})
discussed earlier.

Finally we would like to point out an interesting experimental
realization of QCP in a quantum wire, that of a hollow cylinder
with thin walls\cite{Liu01}. In this case the pair-breaking
parameter that can be easily obtained form the Usadel
equation\cite{Usadel70}, reads
\begin{equation}
\alpha =D\left[ {\frac{{eH}}{{4c}}}\left[ -4n+{\frac{{eH}}{{c}}}%
(r_{1}^{2}+r_{2}^{2})\right] +n^{2}{\frac{{\ln (r_{2}/r_{1})}}{{%
r_{2}^{2}-r_{1}^{2}}}}\right] ,  \label{alpha_cylinder}
\end{equation}
where $r_{1}$ and $r_{2}$ are the inner and outer radii respectively and $n$
is an arbitrary integer. For a thin cylinder ($r_{1}\approx r_{2}\approx r$)
it reduces to
\begin{equation}
\alpha =(D/2r^{2})(\Phi /\Phi _{0}-n)^{2},  \label{cylinder}
\end{equation}
where $\Phi _{0}\equiv \pi c/e$ and $\Phi $ is the flux enclosed
by the cylinder, thereby rendering the classic Little-Park
oscillations\cite {Tinkham} of $T_{c}$ as can be seen from
Eq.(\ref{Tc_vs_H}). Interestingly, for a cylinder with small
enough radius, $r<r_c=\sqrt{D\gamma/4\pi T_{co} } ,$ it is
possible to push the $ T_{c}$ down to zero at magnetic fields
corresponding to half-integer fluxes $\Phi = \Phi _{0}(1/2+n),$ as
was experimentally observed in Ref. \cite{Liu01}. While the
positive fluctuation contribution to conductivity that we would
associate with the classical regime was clearly observed in this
experiment, the conductivity behavior expected for the quantum and
intermediate regimes were not reported so far.

In conclusion, we have investigated the fluctuation correction to
the normal state conductivity in the vicinity of a
parallel-field-induced QCP\ in dirty samples of reduced
dimensions, taking into account both, quantum and thermal
fluctuations within the diagrammatic perturbation theory. Our key
finding is that there are three regimes that show a qualitatively
different behavior ranging from quantum to classical. The
particular temperature and field behavior of the conductivity is
dictated by the choice of path in approaching the QCP while making
the measurement. We have found that for a nanowire (or for a
hollow cylinder) as well as for a thin film, the zero temperature
conductivity correction that also governs the quantum regime, is
{\it negative}, which means that the quantum pairing-fluctuations
increase the resistance to the charge flow. Our findings  imply
that experiments should detect a {\it negative} magnetoresistance
in the quantum regime. For detailed comparison of our results  for
conductivity dependence on temperature and magnetic field with the
experimental data, the weak localization\cite{Gorkov} and
Altshuler-Aronov\cite{Altshuler80} corrections must be subtracted
from the experimental conductivity dependence. Inclusion of these
corrections will not affect the predicted negative sign of the
magnetoresistance at low temperatures  since the weak localization
correction also results in a negative magnetoresistance while the
Altshuler-Aronov correction does not depend on the magnetic field.

\textit{Acknowledgments} We are grateful to Igor Beloborodov,
Vadim Geshkenbein, Thierry Giamarchi, Alex Koshelev, A.I. Larkin,
Revaz Ramazashvilli, Manfred Sigrist and Andrei Varlamov for
useful discussions. N.S\ thanks Argonne National Laboratory and
A.V.L\ thanks Paul Scherrer Institute and ETH--Zurich for their
kind hospitality. This work was supported by the U.S. Department
of Energy, Office of Science, through contract No.
W-31-109-ENG-38.


\begin{thebibliography}{99}
\bibitem{Zeeman} We assume that the orbital effect dominates whereas for
ultrathin wires/films, pair-breaking coming from the Zeeman effect should also be considered.
\bibitem{remark} To avoid a possible confusion, we note that Eqs.(\ref{F1},\ref{F2})
do not assume taking only $\Omega=0$ contribution in the summation
over $\Omega$.
\end{thebibliography}
\end{document}